# TRUST-BASED APPROACHES TOWARDS ENHANCING IOT SECURITY: A SYSTEMATIC LITERATURE REVIEW

Oghenetejiri Okporokpo, Funminiyi Olajide, Nemitari Ajienka and Xiaoqi Ma

[1]Department of Computer Science, Nottingham Trent University, Clifton Lane, Nottingham NG11 8NS

## ABSTRACT

*The continuous rise in the adoption of emerging technologies such as Internet of Things (IoT) by businesses has brought unprecedented opportunities for innovation and growth. However, due to the distinct characteristics of these emerging IoT technologies like real-time data processing, Self-configuration, interoperability, and scalability, they have also introduced some unique cybersecurity challenges, such as malware attacks, advanced persistent threats (APTs), DoS /DDoS (Denial of Service & Distributed Denial of Service attacks) and insider threats. As a result of these challenges, there is an increased need for improved cybersecurity approaches and efficient management solutions to ensure the privacy and security of communication within IoT networks. One proposed security approach is the utilization of trust-based systems and is the focus of this study. This research paper presents a systematic literature review on the Trust-based cybersecurity security approaches for IoT. A total of 23 articles were identified that satisfy the review criteria. We highlighted the common trust-based mitigation techniques in existence for dealing with these threats and grouped them into three major categories, namely: Observation-Based, Knowledge-Based & Cluster-Based systems. Finally, several open issues were highlighted, and future research directions presented.*

## 1. INTRODUCTION

In recent years, the rapid rise in the adoption of Internet of Things (IoT) technology has revolutionized the way businesses operate, establishing new avenues for innovation and growth. The phrase "Internet of Things" was first coined by Kevin Ashton, linking the idea of Radio Frequency Identification (RFID) tags to supply chain management in 1999 [1]. Ever since then, there has been an exceptional level of IoT proliferation with several different applications of the technology across multiple industries including Retail [2] Construction [3], Financial Services [4], Agriculture [5] and Healthcare [6].

Internet of Things technology has consistently been deployed in various forms across multiple industries, such as Smart manufacturing, Smart Power Grid Systems, Smart Cities, Smart Supply Chains etc. resulting in several organizations being able to automate processes, and boost productivity. The total number total of IoT connected devices was estimated to be around 15 billion as of 2020 and is expected to be doubled by the year 2030 [7]. However, the adoption of these IoT technologies has also introduced several unique cybersecurity challenges that cannot be ignored [8].

The ubiquitous nature of IoT devices and their various deployments tend to attract malicious attackers seeking to exploit networks vulnerabilities and to gain unauthorized access to sensitive data and information. This has resulted in an increase in the spate of cyberattacks, ranging from Brute force attacks, Social Engineering attacks, Man in the Middle attacks, Denial of Service (DoS) and Distributed Denial of Service (DDoS) attacks, etc. These attacks not only cause substantial damages to processes and productivity for businesses but have resulted in several privacy and confidentiality concerns [9].

To address these security challenges, it is essential to understand the evolving threats and vulnerabilities associated with IoT technology in terms of confidentiality, integrity, and availability. An enhanced security approach and management solution which offers more secure and reliable network communications is therefore required because such distributed systems often have flexible topologies and have decentralized controls. One such proposed security approach is by means of Trust-based systems [10]. A trust-based system in one which identifies, collates, and makes security decisions based on trust values, and reputation [11]. Specifically, in relation to the application of Trust-based systems to the problem of cyber security in IoT networks, to the best of our knowledge there appears to be very limited Systematic Literature Reviews (SLRs).

In the study by Tyagi et al. [12] the challenges and problems associated with the use of trust management systems for security and privacy in IoT were highlighted. An analysis of trust evaluation and management techniques was done, and these techniques were classified under four major areas: computational, cryptography, and probabilistic, information theory-based and others. There have been some other studies on cyber security techniques for IoT technologies and its broader impact in relation to cyber risk management frameworks. Several review articles [13]-[15] on this have been published. Studies [16]-[20] have reviewed methods for detecting cyberattacks and mitigation of attacks on IoT devices and sensor networks using machine learning and deep learning.

This research paper aims to contribute to this understanding by conducting a systematic literature review on the cybersecurity challenges specific to IoT technology and its various applications of Trust-based systems. By analysing existing studies and research, the review seeks to identify the common types of threats targeting IoT technologies and the corresponding trust-based mitigation techniques employed to counteract them [21].

The contributions of this paper are as follows:

- A systematic literature review (SLR) on the state of the art in literature of trust-based approaches as applied to cybersecurity of IoT.

- A detailed analysis of trust-based systems and techniques are presented in this paper. Also, based on the reviewed literature, these techniques are grouped into three major categories, namely: Observation-Based, Knowledge-Based & Cluster-Based systems. The challenges associated with each of these techniques are also highlighted in this paper.

- A review of the design approaches, key performance metrics for evaluating the efficiency and accuracy of trust-based systems, as well as the advantages, and disadvantages of trust-based cybersecurity techniques.

- Identification of several open issues and challenges for research on trust and reputation in IoT.

## 2. RELATED WORK

Cybersecurity of IoT networks continues to be an interesting area for research and development. As IoT technology evolves, new threats and vulnerabilities continue to emerge. Several reviews have proposed methods to deal with the security challenges common with IoT networks. The approaches deployed vary and authors have focused on different aspects on the security IoT networks by using a systematic review approach. However, the focus of this research work is to review the use of trust-based systems as a means of securing IoT networks. Trust-based management techniques employ a systematic method for effectively managing and ensuring trust within a network [11]. This process usually involves the identification and removal of untrustworthy entities, such as malicious nodes, attacking nodes, malfunctioning nodes, and selfish nodes from the network [22]. A comprehensive summary of the various recent surveys concerning trust-based approaches for IoT security is shown in Table 1 below.

Tan and Azman [23] presented a study of Industrial Internet of Things (IIoT) security architecture and analysed the gap between security requirements of IIoT technologies and their deployed industry countermeasures. The IIoT concept was grouped into a four-layer security architecture based on a defined IIoT CIA+ model. However, in the study, the authors failed to discuss the potential impact of deploying emerging technologies as a means of ensuring the security of IIoT networks. In a study by Kaur et al. [24] attacks, datasets, as well as machine learning algorithms and structures employed in the context of intrusion detection systems for IoT devices were highlighted. A categorisation of attacks targeting IoT devices across various layers and protocols was done. The authors further identified prospective features that can be harnessed by machine learning-based intrusion detection systems to effectively identify diverse attack types. However, there were no specific examples of the results of the identified attacks to add depth and contextualize the outlined efforts and classifications.

Din et al. [25] examined key IoT trust management strategies, emphasizing their advantages and disadvantages while providing descriptions of each approach. However, although the review primarily was centred on IoT, there was a lack of presentation of a classification system, including paper selection criteria and publication year considerations. A study by Shirvani and Masdari [26] focused on exploring trust security in the context of the Internet of Things (IoT) and addressing the challenges associated with managing trust within IoT systems. A comparative analysis of these trust-based schemes was presented incorporating concepts and evaluation metrics drawn from the existing literature. However, the authors did not present any classification and open issues, did not conduct the review systematically and did not describe how the papers were selected or what publication years were considered.

Muzammal et al. [10] investigated the security challenges associated with IoT networks and the Routing Protocol for Low Power and Lossy Networks (RPL). The authors also, explored different approaches for mitigating threats and the importance of trust within IoT. However, the review was not conducted systematically, also, the paper selection processes, and the covered years of the selected papers were not indicated. In the study by Lee, [13] an examination was conducted on IoT cybersecurity technologies and cyber risk management frameworks. Subsequently, a comprehensive four-layer IoT cyber risk management framework was introduced. Additionally, the paper demonstrated the practical application of a linear programming approach for distributing financial resources among multiple IoT cybersecurity projects, offering a proof of concept through an illustrative example. However, the paper selection process, evaluation parameters, applied tools, and open issues were not explored. Also, it was not a systematic literature review. Abdullah et al. [27] reviewed the cybersecurity landscape within the IoT domain, highlighting its security challenges. It also addressed specific security requirements and techniques to mitigate the identified challenges. The authors also explored the potential of blockchain technology as a recommended solution to bolster IoT security. However, the authors neither conducted the review systematically nor defined the process of paper selection as well publication years of reviewed papers.

Table 1. Comparison of other related surveys of Trust-based approaches for IoT security.

| Paper Type | Reference | Main Idea | Publication Year | Paper Selection Process | Open Issues | Classification | Covered Years |
|---|---|---|---|---|---|---|---|
| Survey | [23] | Security of IIoT technologies | 2021 | Not Specified | Highlighted | ✓ | Not Specified |
| Survey | [24] | IoT Security Dataset | 2023 | Not Specified | Not Highlighted | ✓ | Not Specified |
| Survey | [25] | Trust Management for IoT | 2019 | Not Specified | Not Highlighted | X | Not Specified |
| Survey | [26] | Trust-based Security for IoT | 2023 | Not Specified | Highlighted | ✓ | Not Specified |
| Survey | [10] | Trust-based Secure Routing in IoT | 2020 | Not Specified | Not Highlighted | X | Not Specified |
| Survey | [13] | Cybersecurity of IoT | 2020 | Not Specified | Not Highlighted | X | Not Specified |
| Survey | [27] | Cybersecurity of IoT | 2019 | Not Specified | Not Highlighted | X | Not Specified |
| Survey | [28] | Cybersecurity in IoT | 2021 | Specified | Highlighted | ✓ | 2015 - 2021 |
| Survey | [29] | Cyberattack of IoT & IIoT | 2020 | Not Specified | Highlighted | ✓ | Not Specified |
| Survey | [30] | Cyberattack of IoT & IIoT | 2020 | Not Specified | Highlighted | ✓ | Not Specified |
| Survey | [31] | Cyberattack of IoT & IIoT | 2021 | Not Specified | Not Highlighted | X | 2005 - 2017 |
| Survey | [32] | Cyberattack of IoT & IIoT | 2020 | Not Specified | Highlighted | ✓ | Not Specified |
| Survey | [33] | IoT Security Dataset | 2022 | Not Specified | Highlighted | ✓ | Not Specified |
| Survey | [34] | Trust Management for IoT | 2019 | Not Specified | Highlighted | X | Not Specified |
| Survey | [35] | Trust Management for WSNs | 2021 | Not Specified | Highlighted | X | Not Specified |
| Survey | [36] | Trust for IoT Security | 2022 | Not Specified | Not Highlighted | X | Not Specified |
| SLR | [37] | Trust Management in SIoT | 2020 | Specified | Highlighted | ✓ | 2012 - 2022 |
| SLR | [38] | Data Provenance in IoT | 2022 | Specified | Highlighted | ✓ | 2012 - 2022 |
| SLR | [39] | IoT Cyberattack detection | 2022 | Specified | Highlighted | ✓ | 2014 - 2021 |
| SLR | [40] | Cybersecurity of IoT | 2022 | Specified | Highlighted | ✓ | 2016 - 2022 |
| SLR | [12] | Trust Management for IoT | 2023 | Specified | Highlighted | ✓ | 2008 - 2022 |
| SLR | [41] | Trust-based Security for WSNs | 2019 | Specified | Highlighted | ✓ | 2005 - 2017 |
| SLR | [42] | Cybersecurity of IoT | 2022 | Specified | Highlighted | ✓ | 2017 - 2022 |
| **Our Work** | **-** | **Trust-Based Systems and Cyber Security** | **-** | **Specified** | **Highlighted** | **✓** | **2010 - 2023** |

Legend - ✓-Yes, x-No

Ahmad et al. [28] conducted a survey of the enabling cloud-based IoT architecture and classified the cloud security concerns in IoT into four major categories, namely Data, Network and Service, Applications, and People-related security issues. However, the review was primarily focused on cloud network environments. Shah and Sengupta [29] surveyed the various categories of cyberattacks and cyber security vulnerabilities of IoT and IIoT devices. However, the paper failed to present any future research directions or elaborate of research methodology used.

An evaluation of emerging IIoT paradigm was done by Tyagi et al. [12] by identifying specific domains of IIoT adoption, assessing threats and vulnerabilities, and carried out a detailed analysis of existing countermeasures. The authors also highlighted the benefits and challenges of IIoT adoption in industrial sectors with emphasis on the distinctive peculiarities of IIoT deployments. However, most of the security countermeasures presented are designed primarily for consumer IoT and fail to address the highlighted security concerns. A study by Anwar et al. [41] assessed the design and development of trust-based security for wireless sensor networks (WSNs). The results were analysed, with a focus on the designs, applications, and protocols, as well as trust factors. The study suggested that designing the trust management models based on the taxonomy of routing applications and relevant algorithms require further investigations. The study provides a significant contribution to Hybrid and scalable security solutions for trustworthy and secure routing environment for WSNs. However, the scope was very narrow and was focused solely on WSNs.

## 3. METHODOLOGY

A systematic literature review process, based on Barbara Kitchenham's method [43], was carried out with the aim of surveying the existing knowledge about the topic of this article. The SLR methodology employed in this study is depicted in Figure 1. Initially, we meticulously formulated a comprehensive review plan. Subsequently, we identified the research's imperative, established the search and review protocols, and conducted exhaustive searches across various databases as stipulated in the review plan. Following this, we eliminated duplicate articles from the search results and conducted a preliminary review of the remaining articles, resulting in a curated list of potentially valuable articles. These selected articles underwent thorough scrutiny and analysis, culminating in the compilation of a list comprising pertinent articles for this research. Detailed information about the review planning can be found in Section 3, while the outcomes of the search and review process are presented in Section 4.

### 3.1. Review Approach for this Systematic Review Analysis

The specification of the need for this research, along with the delineation of the search and review protocols employed for the Systematic Literature Review, is outlined as follows.

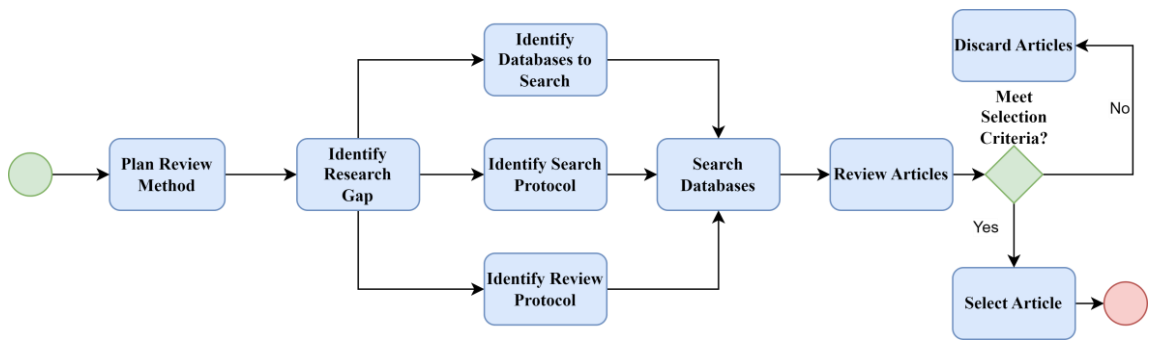

Figure 1. Researcher Approach: A High-level overview of the literature review process

### 3.2. Research Gaps identified

To conduct the literature review, various databases were scrutinized to identify relevant studies published between 2010 and 2023. The selection criteria included papers addressing cybersecurity aspects of IoT, threat analysis, risk assessment, and mitigation strategies. A total of 23 research papers were selected and thoroughly analysed to extract key findings and insights. The analysis of the literature revealed several significant trends and research gaps in the area of trust-based cybersecurity approaches for IoT. The identified trends encompassed authentication mechanisms, encryption techniques, anomaly detection, and intrusion detection systems specific to IoT networks and devices. Moreover, the study uncovered research gaps related to the lack of standardized security frameworks, limited real-world case studies, and the need for more comprehensive threat intelligence and sharing platforms.

To make the work more focused, the following questions were put forward to guide the investigation:

- RQ1 - What is the state of the art of Trust-based systems in addressing cybersecurity needs in IoT environments?

- RQ2 - Which metrics are essential for evaluating and calculating Trust within IoT networks?

- RQ3 - What are the open issues and future challenges of Trust-based systems in the IoT?

The specification of the need for this research, along with the delineation of the search and review protocols employed for the Systematic Literature Review, is outlined as follows.

## 3.3. Resources for Literature Review

The following academic databases were used for this literature review:
SCOPUS, IEEE Xplore, ScienceDirect, Springer. Google Scholar was used as an initial source for general reading material.

## 3.4. Search Protocol

This explains the search protocol applied to the databases. Consequently, the specific keywords (K) utilized for the review, as well as their Combinations (C), have been specified (refer to Table 2). Several general guidelines were established for executing searches across each of the designated resources, including:

- In certain instances, the search terms were input in a stepwise manner, refining the results of preceding searches.

- In cases where search results were restricted in accessibility, alternative avenues, such as authors' personal websites, were explored to locate the documents.

- Consideration was given to the possibility of encountering new terms or concepts that could enhance the discovery of relevant works.

An online reference manager facilitated the documentation of search results and their respective sources. Furthermore, a comprehensive table was employed to log the results of each search, encompassing source details, term combinations, the count of located articles, and the date of each search. For every entry within this table, another table was employed to record references and the evaluation of each reviewed article. Giving a brief description explaining the motive of acceptance or rejection and the acceptance topic to which they belong.

Table 2. Keywords and Combinations used to perform the Systematic Literature Review

| Keywords | K1: IoT | | K4: Attacks | K7: Threats | K10: Systematic |
|---|---|---|---|---|---|
| | K2: Cybersecurity | | K5: Trust | K8: Security | K11: Literature |
| | K3: Cyber attacks | | K6: Trust-based | K9: SLR | K12: Review |

| Combinations | C1: K1 AND K9 |
|---|---|
| | C2: K1 AND K2 |
| | C3: (K2 OR K3) AND K5 AND K9 |
| | C4: K1 AND (K2 OR K3) AND K6 AND K9 |
| | C5: K1 AND K10 AND K11 AND K12 |
| | C6: (K2 OR K3) AND K10 AND K11 AND K12 |
| | C7: K1 AND K2 AND K6 AND (K9 OR (K11 AND K12)) |
| | C8: (K2 OR K3) AND K1O AND K11 AND K12 |
| | C9: K2 AND K9 AND (K5 OR K6) |
| | C10: K1 AND K2 AND K2 AND K4 AND K7 |

## 3.5. Review Protocol

A partial review was conducted to identify potentially valuable research papers. During this review, the abstract of each article was carefully examined. Additionally, in certain cases, the introduction and conclusions of the papers were also reviewed, and in specific instances, relevant sections of the article's body were studied. After thorough examination, each article was evaluated against the criteria specified in the protocol, and a decision was made regarding its inclusion as a potentially valuable resource. To maintain control over the selection process, the tables as described earlier was employed to track accepted and rejected articles. Articles were considered for inclusion if they were related to any of the following Relevant Subject (RS), which are aligned with the research questions outlined above:

- RS1: IoT technology.
- RS2: Cybersecurity approaches for IoT technology.
- RS3: Cybersecurity challenges specific to IoT technology.
- RS4: Trust-Based Systems and frameworks tailored to IoT technology.

Any article which contained the search terms or combinations of them, was initially gathered. An initial review was then carried out by reading through the abstract of each one, followed by a review of their introduction and conclusions. Any article that did not contain relevant information on the subject area at hand, was excluded. Further exclusion criteria used:

- The paper must present empirical data related to IoT and Cybersecurity techniques.

- The paper must be a peer-reviewed and published in a conference proceeding or journal.

- The paper must have been in English language.

Subsequently, a comprehensive assessment of these potential articles was conducted, with each article being scrutinized according to its alignment with the relevant subject areas. For articles falling under RS1 and RS2, a meticulous examination was carried out to identify the unique contribution to the knowledge of IoT technology as well as the limitations and inadequacies of traditional cybersecurity approaches in addressing the evolving security needs of IoT technology. For IoT technology, this entailed recording the reference, the proposed deployment mode, the associated architecture, and a concise description. For Cybersecurity approaches, the information encompassed the reference, the combination of the type and the specific technology involved, including a brief description. As for comparison and selection criteria, the recorded details encompassed the reference, the criteria utilized, and a succinct description. Lastly, for decision outlines, the recorded information included the reference, a brief overview, and an evaluation of their advantages and disadvantages.

In the case of articles classified under RS3 and RS4, an exhaustive review was undertaken to gain a thorough understanding of their proposals and to discern their merits and drawbacks. Emerging cybersecurity challenges specific to IoT technology, and how they differ from those in traditional network environments identified in the articles were reviewed. The primary studies selected were categorized into four acceptance areas, as detailed in Table 3.

Table 3. Accepted Papers grouped into relevant subject areas

| Relevant Subject Area | Number of Accepted Papers |
|---|---|
| RS1: IoT technology. | 8 |
| RS2: Cybersecurity approaches for IoT technology. | 5 |
| RS3: Cybersecurity challenges specific to IoT technology. | 5 |
| RS4: Trust-Based Systems and frameworks tailored to IoT technology. | 5 |
| Total | 23 |

A comprehensive search strategy was implemented, involving 50 distinct combinations of keywords across the specified databases. Figure 2. below illustrates the selection process, at each stage down to the final selection of primary studies. A total of 1,360 articles underwent initial review. To enhance the quality of the results, refinements were introduced in selected databases. In Scopus, the subject area was constrained to Computer Science and Computer Engineering, while Science Direct's content type was restricted to Research Articles and Review Articles. Additionally, in Google Scholar, only review articles where selected, resulting in a total of 125 unique articles. Among the 125 papers reviewed, duplicates were systematically removed, an initial screening identified 46 of these as potentially valuable for the research. Subsequently, a more detailed analysis revealed that 28 of these potential papers were not pertinent to the current research and were therefore excluded. Forward and backward snowballing identified an additional 3 and 2 papers respectively, leaving a final count of 23 papers that were deemed valuable and accepted.

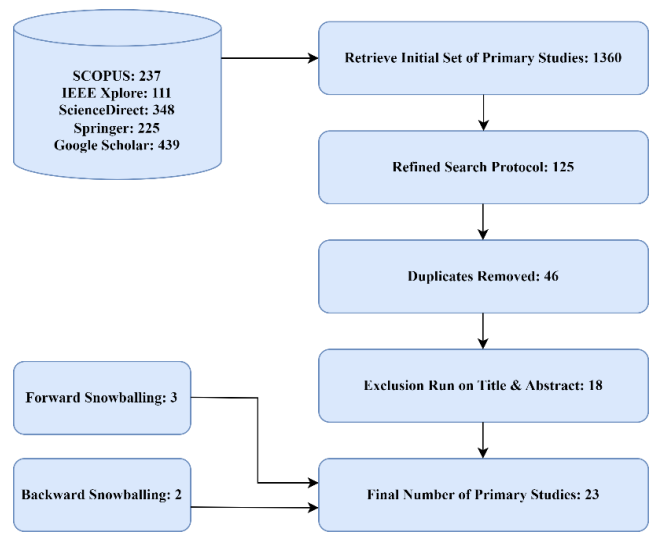

Figure 2. Process of Primary Studies Selection

## 4. RESULTS

Our findings from the review of final selected papers shows that the paradigm of IoT technologies has maintained a consistent amount of interest among researchers in the past 5 years and the idea of deploying trust-based techniques as a means of Cybersecurity is seeing an upward trend. Figure 3. Below shows a chart of the number of primary studies published by year.

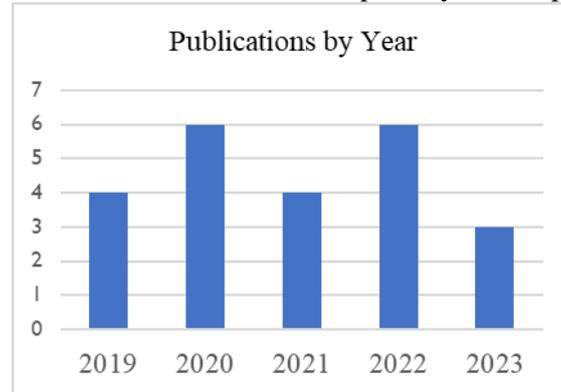

Figure 3. Number of Primary Studies Published by year

To identify the common themes among the selected primary studies, a keyword analysis was conducted across a total of 23 studies. In Table 4 below, we can observe the frequency of specific words across all primary studies. The most frequently appearing keywords in our dataset are "IoT," "IoT Networks," "Trust." , "Cybersecurity," "Cyberattack," and "IoT security. This finding underscores the growing interest in the integration of Trust Management within the context of the Internet of Things (IoT), which we will delve into further in Section 5.

To allow for a simplified classification of the themes of the selected primary studies, they were further grouped into broader categories. Papers with a focus IoT and IIoT were grouped together, papers with a focus on Networking. IoT Networks and Wireless Sensor Networks were clustered together into the Networks category. Papers with a primary focus connected to Trust, Trust Models, Trust Administration, Trust reputation, Trust Values were grouped into the category of Trust Management.

Figure 4. below shows the percentage of themes for the selected primary studies grouped into 6 major categories: IoT, Networks, Trust Management, Cybersecurity, Privacy and Datasets. The themes identified highlight that 45% of studies are focused on IoT, 18% are focused on Networks, while Cybersecurity accounts for 15%. The 3rd most prevalent theme with 12% is Trust Management. Privacy and Datasets account for 5% each.

Table 4. Keyword Counts of Selected Primary Studies

| Keywords | Count |
|---|---|
| Internet of Things (IoT) | 20 |
| IoT Networks | 17 |
| Trust | 15 |
| Cybersecurity | 14 |
| Cyberattack | 12 |
| IoT security | 12 |
| Privacy | 8 |
| Attacks | 7 |
| Trust management techniques | 5 |
| Trust Models | 5 |
| Industrial Internet of Things (IIoT) | 2 |
| Systematic literature review | 4 |
| Reliability | 3 |
| IoT datasets | 3 |
| Trust administration | 3 |
| Wireless sensor networks (WSNs) | 3 |
| Trust values | 2 |
| Protocols | 2 |
| Trust reputation | 2 |
| Intrusion detection system (IDS) | 2 |

Figure 4. below shows the percentage of themes for the selected primary studies grouped into 6 major categories: IoT, Networks, Trust Management, Cybersecurity, Privacy and Datasets. The themes identified highlight that 45% of studies are focused on IoT, 18% are focused on Networks, while Cybersecurity accounts for 15%. The 3rd most prevalent theme with 12% is Trust Management. Privacy and Datasets account for 5% each.

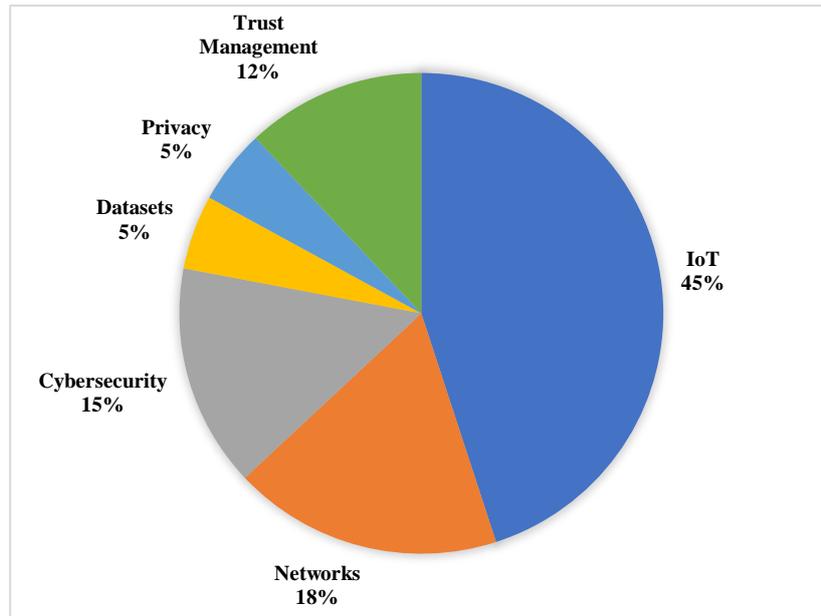

Figure 4. Number of Primary Studies Published by Year

Internet of Things (IoT) being a recent innovative technology has already made some significant contributions to various industries and sectors in a short period of time [32]. The IoT network of physical devices, vehicles, appliances, and other objects embedded with sensors, software, and network connectivity have enhanced communication and data sharing [1]. The implementation of IoT technology has seen an increase in recent years, offering unique opportunities for advances and inventions. However, this rapid proliferation of IoT devices has also brought forth unique cybersecurity challenges that cannot be ignored [44]. As a result of the high-risk potential often involved in the processing of IoT data, the adoption of IoT has faced several challenges and setbacks in various industries [4]-[6]. To maintain data integrity, availability and confidentiality security mechanisms have been put in place [45]. Organizations have continued to deploy numerous solutions to prevent and mitigate cyberattacks and the resultant loses associated with these attacks and in some cases have had to deploy a combination of multiple solutions [46].

There have been different cyber security methods proposed over the years by researchers worldwide, all aimed at either preventing cyberattacks or mitigating the effects of successful attacks [47]. Whilst some of these techniques are fully deployed and in use, there are several others still undergoing research [28].

In the papers reviewed, security and privacy considerations have repeatedly emerged as the pivotal themes, and as a result been the primary focus of most investigations [41]. The privacy and security challenges peculiar to IoT networks oftentimes require adaptive solutions due to the unique attributes of IoT networks/devices, such as their large-scale deployment, diverse ecosystem, and computational capabilities inherent in the design and deployment [40]. Privacy considerations in the design of IoT networks when viewed from a data provenance perspective, has led to scalability issues which introduces potential drawbacks [48]. To address these concerns and maintain privacy, security mechanisms like encryption and authentication have been deployed often [49]. The exploration of trust-based systems as a means of addressing these scalability issues in IoT network designs is a future research endeavour. The examination of cyber security datasets and trust-based systems by investigating novel methods of cloud security, communication and privacy issues is an emerging area of interest [24]. Trust-based and trust-provenance systems, as well as information, network, and database security methods for detecting and responding to cyberattacks in various technologies have been proposed by several researchers [50]-[52]. However, some of the proposed solutions focus on only one type of attack and due to scalability issues and can rarely be integrated into other technologies [51].

## 5. DISCUSSION AND EVALUATION

### 5.1. Basic Concepts

#### 5.1.1. Trust

The concept of trust is multidimensional and has been defined in so many ways across a diverse range of fields from social sciences to computing. The idea of Trust is closely interlinked to several other concepts such as Reputation and Trustworthiness which are fundamental for making decisions in each of these areas. The definitions vary depending on the specific context in which it is being used, for example, in [53], trust is described as a form of confidence, a belief in the expected behaviours or actions of an entity. Reputation on the other hand is defined by Khalid et al. [54] as the opinion an entity has on another's behaviours or actions. Trustworthiness is defined as the cumulative opinion of the behaviour or actions of another entity [55]. The concepts of Trust, Reputation and Trustworthiness have been deployed in various industries for years and in recent times attempts have been made to integrate these concepts when designing IoT networks. In the context of IoT networks, Aldowah et al. [56] define trust as the likelihood of a node adhering to expected behaviour. Similarly, trust can be defined as an agent's ability to provide high-quality services based on a mutually predefined parameter [57]. Another perspective of trust is the degree of belief in a node within a network of nodes [58]. On the other hand, within Wireless Sensor Networks (WSNs), trust is defined as the degree of confidence in the assessment between communicating nodes [10]. In addition, Gautam and Kumar [36] delineates trust as the measure of a nodes capacity to ensure predetermined services. Furthermore, in IoT-based WSNs, trust can be described as the measure of data quality exchanged among sensor nodes [48]. In summary, trust can be characterized as the level of interdependence between a trustor and a trustee based on previously established expectations [59]. Therefore, expectations, ability, capacity, belief, and capability emerge as common elements in the evaluation and establishment of trust.

#### 5.1.2. Node Misbehaviour

In IoT networks, when a node deliberately chooses not to co-operate, or behave in manner expected, they are classified as misbehaving nodes [60]. There are usually two reasons for node misbehaviour; The node may be attempting to conserve power and not spend valuable resources such as memory and CPU cycles for operations that are of no direct benefit to it, this is referred to as a Selfish node or the node may be attempting to cause damage to the network and is known as a Malicious node [52]. Figure 5. below shows the types of node misbehaviours.

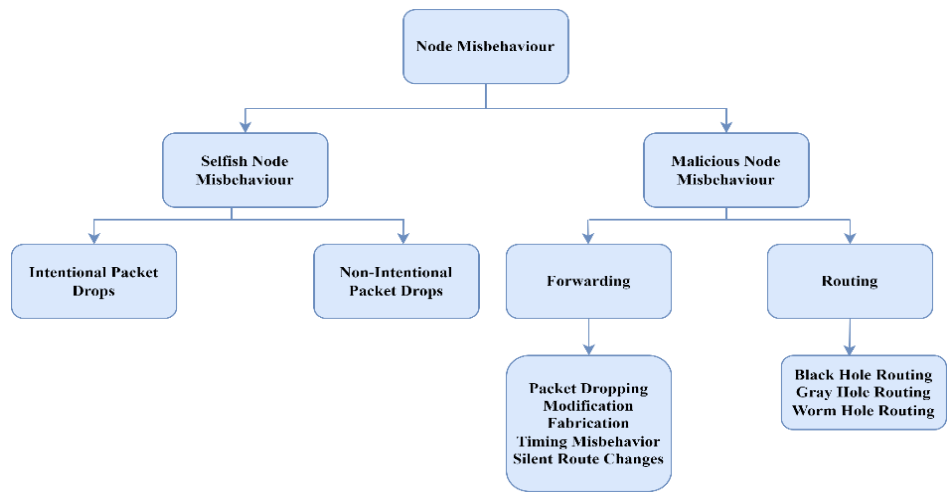

Figure 5. Node Misbehaviours [52]

Selfish nodes tend to exhibit traits that conserve resources and optimize the advantages and an IoT network can still cope when a node exhibits selfish traits. The network can deploy resource incentives to guarantee node co-operation and there is the possibility of anticipating node misbehaviour. There are largely two classes of selfish node misbehaviour; intentional non forwarding/packet drop, and unintentional packet drop [61]. Nodes which possess adequate resources and forwarding capacity, and deliberately choose not to forward packets are wicked-selfish nodes. In the case of unintentional packet drop, occurs when there is a software fault, or a lack of forwarding capacity or memory resources. Malicious nodes introduce large data packets into a network which causes it to be overwhelmed and deplete its resources. Most networks fail to manage malicious node misbehaviour and rely on detecting and removing the node entirely from the network [61]. Malicious node misbehaviour can be largely classed into 2 groups: forwarding and routing [52]. Forwarding malicious nodes usually exhibit traits such as packet dropping, modification, fabrication, timing attacks, and silent route change.

## 5.2. State of The Art of Trust-Based Systems in Addressing Cybersecurity Needs in Iot Networks

### 5.2.1. Trust-Based Systems

With the rapid advances in the realm of IoT networks and Wireless communication networks, the cyber threat landscape has continued to evolve at a similar pace, thus leading to unique threats and challenges. To address this need, a trust-based system is essential, one that can establish the protocols for ensuring safety within an IoT network environment and subsequently detect anomalies. Trust-based systems have been deployed across multiple industries for years. In recent times, attempts have been made to model IoT networks, MANETs and WSNs as reputation-and-trust-based systems [10], [48], [62]. Various types of trust-based systems and techniques for cybersecurity of IoT networks and devices have been proposed by researchers. We have grouped these techniques into three major categories, namely: Observation-Based, Knowledge-Based, Cluster-Based and Hybrid-Based systems. Figure 5. below shows the groups of Trust-based systems.

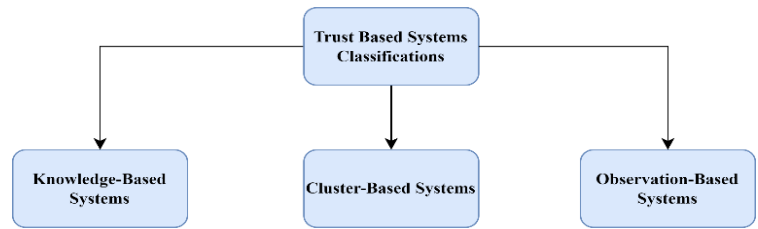

Figure 6. Classifications of Trust-based Systems

### 5.2.2.  Observation-Based Systems

Observation based systems can be broadly categorized in two distinct areas, direct and indirect observation-based systems.

- Direct observation: As the name implies, this type of system relies on direct observation or its own encounters to update reputation and trust [51].

- Indirect Observation: These systems primarily depend shared experiences from other devices or nodes on the network that are passed along depending on the network configuration [50].

Most of the systems proposed employ a combination of both direct and indirect observation to update reputation and trust. This technique allows the system to draw upon the shared encounters of its neighbouring nodes or devices on the network. Wei et al. [51] proposed a trust management scheme for enhancing security in Mobile Ad Hoc Networks. In the proposed system, the trust value is computed using Bayesian inference from an observer node, whilst the indirect observation, is obtained from neighbour nodes and the trust value is calculated using the Dempster-Shafer theory (DST). The module of trust evaluation and update within the trust scheme entity updates reputation based on direct and indirect observation elements and then deploys two techniques; Bayesian inference and DST, to calculate and update the trust values. The trust values are then retained in trust repository module. A secure routing path between sources and destinations nodes can then be established within the networking module based on the trust repository module. Data can be sent via the established secure routing paths by the application module. One benefit of this scheme is that node misbehaviours such as dropping or modifying packets can be quickly detected within the IoT network and the compromised or selfish node excluded from the routing algorithm thereby increasing network throughput. However, one drawback with the proposed scheme is that as the number of nodes in the network increases the total message packets becomes larger the overhead increases and dramatically slows down performance [51].

A trust-based security approach to address Wormhole and Gray hole attacks in mobile ad hoc networks using uncertain reasoning is proposed by Mehta and Parmar [50]. Their method also uses a combination of direct and indirect trust computed based on node observations. In the proposed scheme, each node in RPL (Routing Protocol) network monitors its neighbouring nodes to detect any deviations from the defined protocols. Trust computation is based on trust metrics called as Direct Trust and Indirect Trust. One advantage of the proposed scheme is that it is energy efficient and does not create excessive network overheads. However, it is only effective when the malicious nodes are not colluding. When malicious nodes start collaborating, then they can help prolong the survival time of one another which dramatically reduces the efficiency of the network.

### 5.2.3.  Knowledge-Based Systems

Knowledge-Based systems could be classed as either Symmetric or Asymmetric.

- Symmetric: In Symmetric-based systems, access to information cumulated from direct and indirect observation is accessed by all nodes and devices in the network. Thus, all devices or nodes in the network have the same level of knowledge required for making decisions [52].

- Asymmetric: In Asymmetric-based systems, not all devices or nodes in the network have knowledge of all information.

A trust-based approach for anonymous communication using asymmetric cryptography scheme is proposed by Wenjia and Song [63]. The proposed scheme is made up of two phases, namely data analysis and trust management. It can detect and cope with malicious attacks and evaluate the trustworthiness of both data and mobile nodes in VANETs. However, one major drawback of such Asymmetric-based systems is that since not all nodes on the network have access to the same amount of information, as the node density increases, the scheme introduces additional communication overhead.

### 5.2.4.  Cluster-Based Systems

Cluster-Based systems usually have a central device or node which carries out all the trust computations and maintains a repository of the status of all the nodes in the network. Improved security or efficiency is a major achievement of Cluster-based systems. Some class of cluster-based systems have all devices or nodes in the network and maintain a repository of residual information of the status of the other nodes and devices in the network. Researchers [64] proposed a trust-based Information sharing schemes for distributed collaborative networks. Their approach uses a central trust authority based on the recently proposed identity-based broadcast encryption (IBBE) technique.

A cluster-based trust management model for centralized cognitive radio networks is proposed by Qingqi et al [65]. The model can detect the malicious behaviour untrustworthy nodes. As shown in Figure 10 below in the central structure, the primary users and the second users distribute in the same geographic area. The primary base station (PBS) controls the primary users. The cognitive base station (CBS) controls the second users. One advantage of the system is that it can detect malicious node behaviour in the network, and protecting genuine second users from soft cyberattacks and collusion attacks to provide reliable security assurance for dynamic spectrum access. However, the major drawback is that the trust mechanism is unique to the cognitive radio networks.

### 5.3.  State of the art of Trust-based systems in addressing cybersecurity needs in IoT Networks

Trust metrics are essential for identifying, calculating, and evaluating trust values and trust properties in an IoT network. The metrics deployed are usually largely dependent on several factors including the IoT network in question, the trust management technique as well as the evaluation parameters available. Trust properties is a commonly identified metric used for trust formation in the reviewed literature [52].

### 5.3.1.  Properties of Trust

Trust-based systems are primarily comprised of three distinct properties. Durable nodes/devices that cumulate a repository of protocols for future communication, compilation, and dissemination of information regarding ongoing communications and ensuring its availability for future reference and deployment of a propagation mechanism to aid the dissemination of trust information to peer nodes/devices on the network. According to researchers [64], the main constituents the trust metric are:

Knowledge-Based systems could be classed as either Symmetric or Asymmetric.

- Asymmetry: Trust is not symmetric in nature, meaning that, because node/device 1 trusts node/device 2, does not inevitably mean that node/device 2 trusts device 1

- Transitivity: Trust is transitive in nature, meaning that, if node/device 1 trusts node/device 2 and node/device 2 trusts node/device 3, then node/device 1 trusts node/device 3.

- Reflexivity: Trust is reflexive in nature, meaning that a node/devices' default setting is to trust itself.

### 5.3.2.  Trust-Based System Components

There are several identified components of a trust-based system from the reviewed literature. One common approach is to classify trust-based models into five distinct areas: trust composition, trust aggregation, trust propagation, trust updation, and trust formation [66]. Figure 6 below shows the identified trust-based system components.

### 5.3.2.1. Trust Composition

Trust composition includes the necessary elements required for effective trust computation [64]. These elements comprise of quality of service (QoS) trust and social trust. QoS trust refers to the expectation that a node/device in an IoT network will provide a guaranteed quality of service or act in a manner as previously agreed upon based on pre-defined parameters such as execution time, availability, completion rate, execution time, turnaround time, and universal accessibility [67].

Social trust is particularly prominent is Social Internet of Things (SIoT)  and refers to the relationship that exists between nodes/devices in IoT networks, and the owner/vendor of these networks. Social trust is usually used to evaluate IoT nodes/devices based on several parameters such as security, reliability, and connectivity [64].

### 5.3.2.2.  Trust Aggregation

Trust aggregation involves the collation of reputation information gathered from direct-observations or indirect peer observation [64]. Prominent techniques for trust aggregation examined in the existing literature encompass fuzzy logic [68], regression analysis [69], weighted summation [35], Bayesian inference  [70], and belief theory [71].

### 5.3.2.3. Trust Propagation

Trust propagation is concerned with disseminating first-hand information gathered by nodes to node/device peers on the network [66]. The two common trust propagation schemes are distributed and centralized. In a centralized system, a single node/device maintains a central repository of the reputation of all network nodes/devices. Alshehri and Hussain [72] proposed the CTM-IoT mechanism which uses a Super Node (SN) as the centralized trust manager node. In their approach, the IoT network is divided into clusters to achieve trustworthy communication between nodes with each cluster having a local trust manager called a Master Node (MN). One drawback of this mechanism is that it can create security and information bottlenecks if the central node is attacked. In a distributed system, each node/device in the system independently stores a repository of reputation information for peer nodes/devices. In this type of reputation system, challenges may arise regarding the consistency of reputation values across various nodes, potentially leading to a lack of coherence. A lightweight mechanism for mobile devices that effectively propagates trust and is distributed was designed by Quercia et al. [73]. This mechanism uses a graph-based learning technique where nodes/devices are either rated or unrated, and those nodes are then connected to each other if they are related. The technique considers two nodes to be related when they possess identical ratings. Whilst distributed systems solve the problems associated with centralized systems, they usually have large overheads often resulting in issues with scalability.

### 5.3.2.4.  Trust Updation

Trust updating deals with updating the trust value of each node/device on the IoT network [66]. Some IoT systems are designed to update trust at periodical intervals (time based) [74], whilst some are designed to update trust information based on discrete events (event-based) [22].

### 5.3.2.5.  Trust Formation

Trust formation refers to the process of decision making, and trust assessment based on various trust attributes [66]. In the reviewed literature, trust formation techniques are usually either single-trust and multi-trust based in IoT networks [29], [39], [41], [42], [66]. Trust-based decisions depend on reputation information provided by the aggregation component. The fundamental decision is binary, deciding on which node/device to trust and which node/device not to trust. This decision may result in actions such as cooperate/don't cooperate, forward/don't forward, and so on, dependent upon the specific agreed upon matrices by the system.

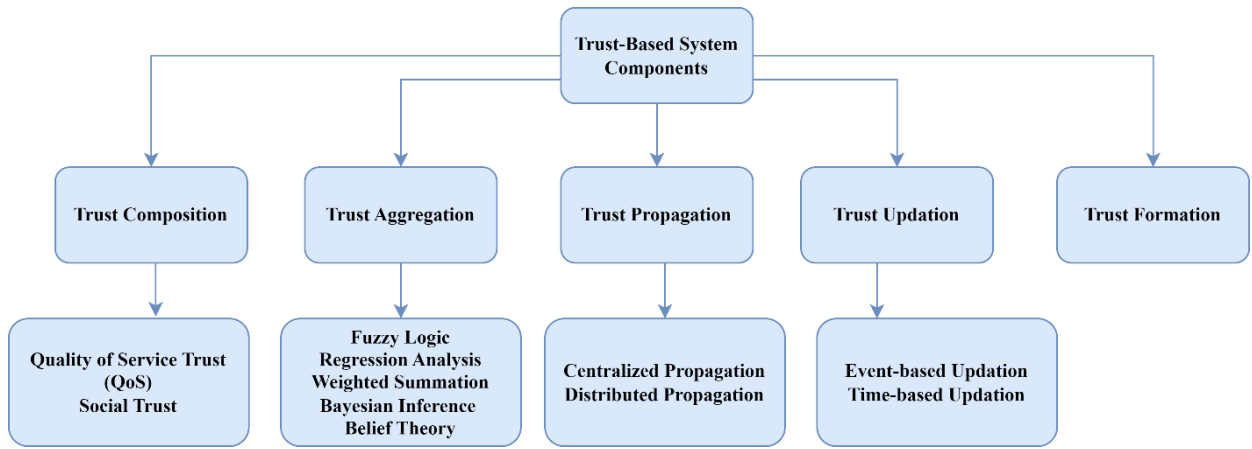

Figure 7. Trust-based Systems components and metrics

Some other metrics used for evaluating trust-based systems include.

**Reliability:** This involves ensuring that the IoT network functions with minimal errors or interruptions within a defined timeframe [72]. Implementing trust-based techniques requires the deployment of nodes that can carry out tasks that enhance the overall reliability of the network with availability within specified parameters.

**Efficiency:** This involves measuring the accuracy of the trust-based system in calculating trust values and node reputation within pre-defined parameters [70].

**Integrity:** This ensures that the node reputation information and its content are secure and unaltered during communication between nodes in the IoT network [68].

**Scalability:** This refers to the capacity to which a trust-based system can be adapted to meet evolving conditions as the number of devices, nodes, or network needs changes [67].

**Availability:** This guarantees that even when under attack the IoT network continues to proffer services [75]. Most trust-based systems achieve this by node segregation or clustering methods [51], [76].

**Serviceability:** This guarantees that IoT network is easily modified or updated to accommodate necessary changes when an attack has been detected or new network vulnerabilities are identified [75].

## 5.4. Open issues and Future Challenges of Trust-Based Systems in the Iot

From the reviewed literature we have identified several open issues and challenges for research on trust-based techniques in IoT.

- **Scalable Trust-based System for IoT:** With the projected increase in the number of IoT connected devices, there is a need to design scalable trust-based IoT solutions that can be easily adapted to accommodate the growing network demands. Most of the reviewed literature rarely mention trust-management techniques that can effectively resolve the scalability issues.

- **Trust-based DDoS Attack Detection System:** From the review literature, one of the most common attacks targeting IoT devices and networks are DDoS attacks. There seems to be a limited number of proposed Trust-based DDoS detection system. This would be an interesting area for further research.

- **AI and Machine learning as real-time Detection techniques for IoT Networks:** Real-time detection techniques such as AI technology and machine learning can be adopted for enhancing the security in IoT networks. This is another interesting area that needs to be explored.

- **Blockchain for IoT Network Security:** From the reviewed literature very, few discussions have been had on the use of blockchain for the security of IoT networks. More work needs to be done in exploring the design of lightweight blockchain-based solutions for resource-constrained IoT networks.

- **Edge and Cloud Infrastructure as a Service (IaaS) for IoT Networks:** The design of a scalable, easily adaptable cloud/edge computing infrastructure as a service solution for IoT networks is another area that requires further research.

## CONCLUSIONS AND RECOMMENDATIONS

We conducted a systematic literature review on systematic literature review (SLR) on the state of the art in literature of trust-based approaches as applied to cybersecurity of IoT. Based on the background studies, three research questions were created and a methodology was formulated to assist in searching the relevant databases to answer these questions. From the reviewed literature, trust-based cybersecurity for IoT is an emerging area of interest with an increase in the number of recent studies. Trust-based systems that can detect and prevent cyberattacks can significantly improve privacy and security within IoT networks. The need for ongoing research and development in this area is therefore imperative. As IoT technology continues to evolve, so will the threats and vulnerabilities. Therefore, it is necessary to explore further developments and enhancement of existing trust-based techniques for IoT security. This includes investigating methods for anomaly detection and real-time monitoring aimed at proactively identifying and responding to potential cybersecurity incidents. This research has focused on studies related to IoT and the use of Trust-based techniques and their various applications. A review of the design approaches, and the key performance metrics for evaluating the efficiency of trust-based cybersecurity techniques was done. Finally, several open issues were identified and further research directions on trust-based techniques in IoT networks were presented.

## ACKNOWLEDGEMENTS

The authors would like to acknowledge the support of the Nottingham Trent University (NTU) for a fully funded studentship. The authors also declare that they have no known competing financial interests or personal relationships that could have appeared to influence the work reported in this paper.

## REFERENCES

[1]    K. Ashton, "That 'internet of things' thing," RFID Journal, vol. 22, (7), pp. 97-114, 2009.
[2]    M. S. Hossain et al, "Role of Internet of Things (IoT) in retail business and enabling smart retailing experiences," Asian Business Review, vol. 11, (2), pp. 75-80, 2021.

[3] Y. Gamil et al, "Internet of things in construction industry revolution 4.0: Recent trends and challenges in the Malaysian context," Journal of Engineering, Design and Technology, vol. 18, (5), pp. 1091-1102, 2020.

[4] F. Khanboubi, A. Boulmakoul and M. Tabaa, "Impact of digital trends using IoT on banking processes," Procedia Computer Science, vol. 151, pp. 77-84, 2019.

[5] J. Xu, B. Gu and G. Tian, "Review of agricultural IoT technology," Artificial Intelligence in Agriculture, vol. 6, pp. 10-22, 2022.

[6] Y. A. Qadri et al, "The future of healthcare internet of things: a survey of emerging technologies," IEEE Communications Surveys & Tutorials, vol. 22, (2), pp. 1121-1167, 2020.

[7] L. S. Vailshery, "Number of internet of things (IoT) connected devices worldwide from 2019 to 2023, with forecasts from 2022 to 2030," Transforma Insights; Exploding Topics, Worldwide, "July 23, ". 2023.

[8] K. Kimani, V. Oduol and K. Langat, "Cyber security challenges for IoT-based smart grid networks," International Journal of Critical Infrastructure Protection, vol. 25, pp. 36-49, 2019. Available: https://www.sciencedirect.com/science/article/pii/S1874548217301622. DOI: 10.1016/j.ijcip.2019.01.001.

[9] S. H. Mekala et al, "Cybersecurity for Industrial IoT (IIoT): Threats, countermeasures, challenges and future directions," Comput. Commun., vol. 208, pp. 294-320, 2023. Available: https://www.sciencedirect.com/science/article/pii/S0140366423002189. DOI: 10.1016/j.comcom.2023.06.020.

[10] S. M. Muzammal, R. K. Murugesan and N. Z. Jhanjhi, "A comprehensive review on secure routing in internet of things: Mitigation methods and trust-based approaches," IEEE Internet of Things Journal, vol. 8, (6), pp. 4186-4210, 2020.

[11] U. Rahamathullah and E. Karthikeyan, "A lightweight trust-based system to ensure security on the Internet of Battlefield Things (IoBT) environment," International Journal of System Assurance Engineering and Management, pp. 1-13, 2021.

[12] H. Tyagi, R. Kumar and S. K. Pandey, "A detailed study on trust management techniques for security and privacy in IoT: Challenges, trends, and research directions," High-Confidence Computing, pp. 100127, 2023.

[13] I. Lee, "Internet of Things (IoT) cybersecurity: Literature review and IoT cyber risk management," Future Internet, vol. 12, (9), pp. 157, 2020.

[14] K. Kandasamy et al, "IoT cyber risk: A holistic analysis of cyber risk assessment frameworks, risk vectors, and risk ranking process," EURASIP Journal on Information Security, vol. 2020, (1), pp. 1-18, 2020.

[15] N. M. Karie et al, "A review of security standards and frameworks for IoT-based smart environments," IEEE Access, vol. 9, pp. 121975-121995, 2021.

[16] M. Abdullahi et al, "Detecting cybersecurity attacks in internet of things using artificial intelligence methods: A systematic literature review," Electronics, vol. 11, (2), pp. 198, 2022.

[17] U. Inayat et al, "Learning-based methods for cyber attacks detection in IoT systems: A survey on methods, analysis, and future prospects," Electronics, vol. 11, (9), pp. 1502, 2022.

[18] M. Kuzlu, C. Fair and O. Guler, "Role of artificial intelligence in the Internet of Things (IoT) cybersecurity," Discover Internet of Things, vol. 1, pp. 1-14, 2021.

[19] M. b. Mohamad Noor and W. H. Hassan, "Current research on Internet of Things (IoT) security: A survey," Computer Networks, vol. 148, pp. 283-294, 2019. Available: https://www.sciencedirect.com/science/article/pii/S1389128618307035. DOI: 10.1016/j.comnet.2018.11.025.

[20] S. N. Matheu-García et al, "Risk-based automated assessment and testing for the cybersecurity certification and labelling of IoT devices," Computer Standards & Interfaces, vol. 62, pp. 64-83, 2019. Available: https://www.sciencedirect.com/science/article/pii/S0920548918301375. DOI: 10.1016/j.csi.2018.08.003.

[21] R. Kumar et al, "What changed in the cyber-security after COVID-19?" Comput. Secur., vol. 120, pp. 102821, 2022. Available: https://www.sciencedirect.com/science/article/pii/S0167404822002152. DOI: 10.1016/j.cose.2022.102821.

[22] D. Díaz López et al, "Shielding IoT against cyber-attacks: An event-based approach using SIEM," Wireless Communications and Mobile Computing, vol. 2018, 2018.

[23] S. F. Tan and A. Samsudin, "Recent technologies, security countermeasure and ongoing challenges of Industrial Internet of Things (IIoT): A survey," Sensors, vol. 21, (19), pp. 6647, 2021.

[24]  B. Kaur et al, "Internet of Things (IoT) security dataset evolution: Challenges and future directions," Internet of Things, vol. 22, pp. 100780, 2023. Available: https://www.sciencedirect.com/science/article/pii/S2542660523001038. DOI: 10.1016/j.iot.2023.100780.

[25]  I. U. Din et al, "Trust management techniques for the Internet of Things: A survey," IEEE Access, vol. 7, pp. 29763-29787, 2018.

[26]  M. Hosseini Shirvani and M. Masdari, "A survey study on trust-based security in Internet of Things: Challenges and issues," Internet of Things, vol. 21, pp. 100640, 2023. Available: https://www.sciencedirect.com/science/article/pii/S2542660522001214. DOI: 10.1016/j.iot.2022.100640.

[27]  A. Abdullah et al, "CyberSecurity: A review of internet of things (IoT) security issues, challenges and techniques," in 2019 2nd International Conference on Computer Applications & Information Security (ICCAIS), 2019.

[28]  W. Ahmad et al, "Cyber security in IoT-based cloud computing: A comprehensive survey," Electronics, vol. 11, (1), pp. 16, 2021.

[29]  Y. Shah and S. Sengupta, "A survey on classification of cyber-attacks on IoT and IIoT devices," in 2020 11th IEEE Annual Ubiquitous Computing, Electronics & Mobile Communication Conference (UEMCON), 2020.

[30]  R. W. Anwar, A. Zainal and S. Iqbal, "Systematic literature review on designing trust-based security for WSNs," Indonesian Journal of Electrical Engineering and Computer Science, vol. 14, (3), pp. 1395-1404, 2019.

[31]  M. K. Kagita et al, "A review on cyber crimes on the Internet of Things," Deep Learning for Security and Privacy Preservation in IoT, pp. 83-98, 2022.

[32]  L. L. Dhirani, E. Armstrong and T. Newe, "Industrial IoT, cyber threats, and standards landscape: Evaluation and roadmap," Sensors, vol. 21, (11), pp. 3901, 2021.

[33]  J. Sengupta, S. Ruj and S. D. Bit, "A comprehensive survey on attacks, security issues and blockchain solutions for IoT and IIoT," Journal of Network and Computer Applications, vol. 149, pp. 102481, 2020.

[34]  H. Ahmetoglu and R. Das, "A comprehensive review on detection of cyber-attacks: Data sets, methods, challenges, and future research directions," Internet of Things, pp. 100615, 2022.

[35]  A. Abdlrazaq and S. Varol, "A trust management model for IoT," in 2019 7th International Symposium on Digital Forensics and Security (ISDFS), 2019.

[36]  A. K. Gautam and R. Kumar, "A comprehensive study on key management, authentication and trust management techniques in wireless sensor networks," SN Applied Sciences, vol. 3, (1), pp. 50, 2021.

[37]  J. Ortiz and M. M. Chowdhury, "Computational trust for securing IoT," in 2022 IEEE International Conference on Electro Information Technology (eIT), 2022.

[38]  R. K. Chahal, N. Kumar and S. Batra, "Trust management in social Internet of Things: A taxonomy, open issues, and challenges," Comput. Commun., vol. 150, pp. 13-46, 2020.

[39]  E. Gultekin and M. S. Aktas, "Systematic literature review on data provenance in internet of things," in International Conference on Computational Science and its Applications, 2022.

[40]  A. Sgueglia et al, "A systematic literature review of IoT time series anomaly detection solutions," Future Generation Comput. Syst., vol. 134, pp. 170-186, 2022.

[41]  E. Altulaihan, M. A. Almaiah and A. Aljughaiman, "Cybersecurity threats, countermeasures and mitigation techniques on the IoT: future research directions," Electronics, vol. 11, (20), pp. 3330, 2022.

[42]  I. S. Utomo et al, "A systematic literature review of privacy, security, and challenges on applying IoT to create smart home," in 2022 International Conference on Electrical and Information Technology (IEIT), 2022.

[43]  B. Kitchenham and P. Brereton, "A systematic review of systematic review process research in software engineering," Information and Software Technology, vol. 55, (12), pp. 2049-2075, 2013. Available: https://www.sciencedirect.com/science/article/pii/S0950584913001560. DOI: 10.1016/j.infsof.2013.07.010.

[44]  M. M. Shurman, R. M. Khrais and A. A. Yateem, "IoT denial-of-service attack detection and prevention using hybrid IDS," in 2019 International Arab Conference on Information Technology (ACIT), 2019, .

[45]  M. Ghiasi et al, "A comprehensive review of cyber-attacks and defense mechanisms for improving security in smart grid energy systems: Past, present and future," Electr. Power Syst. Res., vol. 215,

pp. 108975, 2023. Available: https://www.sciencedirect.com/science/article/pii/S0378779622010240. DOI: 10.1016/j.epsr.2022.108975.

[46] M. Humayun et al, "Cyber Security Threats and Vulnerabilities: A Systematic Mapping Study," Arabian Journal for Science and Engineering, vol. 45, (4), pp. 3171-3189, 2020. Available: https://doi.org/10.1007/s13369-019-04319-2. DOI: 10.1007/s13369-019-04319-2.

[47] S. Ambore et al, "A resilient cybersecurity framework for Mobile Financial Services (MFS)," Journal of Cyber Security Technology, vol. 1, (3-4), pp. 202-224, 2017.

[48] G. Jayakumar and G. Gopinath, "Ad hoc mobile wireless networks routing protocols–a review," Journal of Computer Science, vol. 3, (8), pp. 574-582, 2007.

[49] M. Papathanasaki, L. Maglaras and N. Ayres, "Modern Authentication Methods: A Comprehensive Survey," IntechOpen Journals, 2022.

[50] R. Mehta and M. M. Parmar, "Trust based mechanism for securing iot routing protocol rpl against wormhole &grayhole attacks," in 2018 3rd International Conference for Convergence in Technology (I2CT), 2018, .

[51] Z. Wei et al, "Security enhancements for mobile ad hoc networks with trust management using uncertain reasoning," IEEE Transactions on Vehicular Technology, vol. 63, (9), pp. 4647-4658, 2014.

[52] A. Srinivasan et al, "Reputation-and-trust-based systems for ad hoc networks," Algorithms and Protocols for Wireless and Mobile Ad Hoc Networks, vol. 375, pp. 375-404, 2009.

[53] A. J. I. Jones, "On the concept of trust," Decis. Support Syst., vol. 33, (3), pp. 225-232, 2002. Available: https://www.sciencedirect.com/science/article/pii/S0167923602000131. DOI: 10.1016/S0167-9236(02)00013-1.

[54] O. Khalid et al, "Comparative study of trust and reputation systems for wireless sensor networks," Security and Communication Networks, vol. 6, (6), pp. 669-688, 2013.

[55] A. Kumar, V. Capraro and M. Perc, "The evolution of trust and trustworthiness," Journal of the Royal Society Interface, vol. 17, (169), pp. 20200491, 2020.

[56] H. Aldowah, S. Ul Rehman and I. Umar, "Trust in iot systems: a vision on the current issues, challenges, and recommended solutions," Advances on Smart and Soft Computing: Proceedings of ICACIn 2020, pp. 329-339, 2021.

[57] R. Shaikh and M. Sasikumar, "Trust model for measuring security strength of cloud computing service," Procedia Computer Science, vol. 45, pp. 380-389, 2015.

[58] T. Lynn, L. van der Werff and G. Fox, "Understanding trust and cloud computing: An integrated framework for assurance and accountability in the cloud," Data Privacy and Trust in Cloud Computing: Building Trust in the Cloud through Assurance and Accountability, pp. 1-20, 2021.

[59] D. M. Rousseau et al, "Not so different after all: A cross-discipline view of trust," Academy of Management Review, vol. 23, (3), pp. 393-404, 1998.

[60] K. Balakrishnan, J. Deng and V. K. Varshney, "TWOACK: Preventing selfishness in mobile ad hoc networks," in IEEE Wireless Communications and Networking Conference, 2005, 2005, .

[61] W. Khalid et al, "A taxonomy on misbehaving nodes in delay tolerant networks," Comput. Secur., vol. 77, pp. 442-471, 2018.

[62] H. Maddar, W. Kammoun and H. Youssef, "Effective distributed trust management model for Internet of Things," Procedia Computer Science, vol. 126, pp. 321-334, 2018.

[63] W. Li and H. Song, "ART: An attack-resistant trust management scheme for securing vehicular ad hoc networks," IEEE Transactions on Intelligent Transportation Systems, vol. 17, (4), pp. 960-969, 2015.

[64] H. Lin et al, "Efficient trust based information sharing schemes over distributed collaborative networks," IEEE J. Select. Areas Commun., vol. 31, (9), pp. 279-290, 2013.

[65] Q. Pei, R. Liang and H. Li, "A trust management model in centralized cognitive radio networks," in 2011 International Conference on Cyber-Enabled Distributed Computing and Knowledge Discovery, 2011, .

[66] J. Guo and R. Chen, "A classification of trust computation models for service-oriented internet of things systems," in 2015 IEEE International Conference on Services Computing, 2015, .

[67] P. Kumar, S. Vinodh Kumar and L. Priya, "QoS-based classical trust management system for the evaluation of the trustworthiness of a cloud resource," in Inventive Systems and Control: Proceedings of ICISC 2022Anonymous 2022, .

[68] D. Chen et al, "TRM-IoT: A trust management model based on fuzzy reputation for internet of things," Computer Science and Information Systems, vol. 8, (4), pp. 1207-1228, 2011.

[69]  F. A. M. Solomon, G. W. Sathianesan and R. Ramesh, "Logistic Regression Trust-A Trust Model for Internet-of-Things Using Regression Analysis." Computer Systems Science & Engineering, vol. 44, (2), 2023.

[70]  C. V. L. Mendoza and J. H. Kleinschmidt, "A distributed trust management mechanism for the Internet of things using a multi-service approach," Wireless Personal Communications, vol. 103, pp. 2501-2513, 2018.

[71]  C. Esposito et al, "Robust decentralised trust management for the internet of things by using game theory," Information Processing & Management, vol. 57, (6), pp. 102308, 2020.

[72]  M. D. Alshehri and F. K. Hussain, "A centralized trust management mechanism for the internet of things (CTM-IoT)," in Advances on Broad-Band Wireless Computing, Communication and Applications: Proceedings of the 12th International Conference on Broad-Band Wireless Computing, Communication and Applications (BWCCA-2017), 2018, .

[73]  D. Quercia, S. Hailes and L. Capra, "Lightweight distributed trust propagation," in Seventh IEEE International Conference on Data Mining (ICDM 2007), 2007.

[74]  S. Alam, M. M. Chowdhury and J. Noll, "Senaas: An event-driven sensor virtualization approach for internet of things cloud," in 2010 IEEE International Conference on Networked Embedded Systems for Enterprise Applications, 2010, .

[75]  R. Dalal, M. Khari and Y. Singh, "Survey of trust schemes on ad-hoc network," in Advances in Computer Science and Information Technology. Networks and Communications: Second International Conference, CCSIT 2012, Bangalore, India, January 2-4, 2012. Proceedings, Part I 2, 2012, .

[76]  N. Khanna and M. Sachdeva, "Study of trust-based mechanism and its component model in MANET: Current research state, issues, and future recommendation," International Journal of Communication Systems, vol. 32, (12), pp. e4012, 2019.

## AUTHORS

Oghenetejiri Okporokpo is a PhD candidate of computer science at Nottingham Trent University UK.

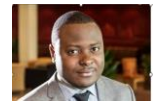

Dr Funminiyi Olajide, is a Senior Lecturer in Cyber Security and Forensics. He is part of the project supervision team.

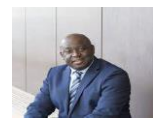

Dr Nemitari Ajienka is a Senior Lecturer at Nottingham Trent University, UK. He is part of the project supervision team.

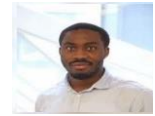

Dr Xiaoqi Ma is currently a Senior Lecturer in Department of Computer Science at Nottingham Trent University, UK. He is part of the project supervision team.

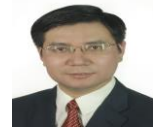